\documentclass[preprint,floats,aps]{revtex4}
\usepackage{dcolumn}
\usepackage{bm}
\usepackage{graphicx}
\begin{document}
\topmargin=-0.3cm
\title{Mean Field Effects  In The Quark-Gluon Plasma}
\author{ Zhi Guang Tan$^{1-2}$, 
and A. Bonasera$^{1,3}$
 \footnote{Email: bonasera@lns.infn.it}}
\affiliation{
$^1$  Laboratorio Nazionale del Sud, Istituto Nazionale Di Fisica Nucleare,
      Via S. Sofia 44, I-95123 Catania, Italy \\
$^2$  Institute of Particle Physics, Huazhong Normal University, Wuhan,
      430079 China\\
$^3$ Libera Universita'  Kore, via della Cittadella 1, I-94100 Enna, Italy      
}
\begin{abstract}
A transport model  based on the mean free path approach for an interacting meson system at finite temperatures is discussed.  A transition to a quark gluon plasma  is included within the framework of the  MIT bag model.  The results obtained compare very well with Lattice QCD calculations when we include a mean field in the QGP phase due to the
Debye color screening.   In particular the cross over to the QGP at about 175 MeV temperature  is nicely reproduced.   We also discuss a possible scenario  for hadronization which is especially important for temperatures below the QGP phase transition. \\
\noindent{PACS numbers: 12.38.Mh, 12.39.Ba}
\end{abstract}
\maketitle

Numerical calculations of a system of hadrons and the transition to the quark-gluon plasma (QGP) at 
finite temperatures and zero baryonic density within a Lattice QCD (LQCD) approach are feasible nowadays  thanks to very performing computers\cite{karsch}.  Up to date LQCD results suggest that there is a cross over from a meson system to a QGP at a temperature of about 175 MeV \cite{karsch}.  These features could be experimentally confirmed or rejected  in 
relativistic heavy ion collisions (RHIC).  In particular high temperatures and very small baryonic densities can be obtained at the very high beam energies reached at the Brookhaven National Laboratory (BNL) and soon at the Large Hadron Collider (LHC) at CERN.  Because of the high complexity of
the problem, models are needed that take into account the relevant physics at such high energy densities, plus the possibility that the system is out of equilibrium during the collisions.  Of course exact  microscopic  simulations for out of equilibrium-finite systems are out of reach at present.  On the other hand transport approaches  \cite{aich,gei} have been very useful in the past in describing many features of lower energies heavy ion collisions.  Generalizations to relativistic energies of low energy heavy ion collisions \cite{cug,bert,aich,sa1} (known as  Boltzmann Uehling Uehlenbeck (BUU), Vlasov(VUU)/ Landau (LV)) have been proposed.
The method we discuss in
this work is known as Boltzmann Nordheim Vlasov (BNV) approach at low energies\cite{aldo,daimei,tan}.  It
is based on the concept of the mean free path approach\cite{aldo}.
In a previous  paper we studied the equilibrium features of our model, i.e. the equation of state (EOS) effectively implemented in the model.
To this purpose, in our calculations  pions are enclosed in a box with periodic boundary conditions to simulate an infinite system.  The pions can collide elastically or inelastically according to the elementary cross sections which are parametrized from available data.  If we restrict our calculations to an hadron system only and we include all the measured resonances ($\rho$, $\omega$ etc..) we obtain the so called Hagedorn limiting temperature, i.e. when we increase the energy density of the system we do not obtain an increase of the temperature as well because  higher mass resonances are excited thus reducing the available kinetic energy \cite{tan}.  
  We can easily include the possibility of a QGP using the bag model \cite{las,wong}. In fact, for each elementary hadron-hadron collision, we can calculate the local energy density and the pressure.  If such quantities overcome the Bag pressure and energy density, then $n_q=n_{\bar q}$ quarks and antiquarks and $n_g$ gluons are created.  The number of quarks and gluons are calculated assuming local thermal equilibrium.  In this way we can simulate a hadron gas and its transition to the QGP.  In \cite{tan},
  we discussed  the cases of $N_f=0,2$, where $N_f$ is the number of flavors.  In this paper we generalize our results to 3 flavors, and compare to LQCD results.  We also discuss the possibility the quarks could recombine and gluons decay into two quarks 
  during the dynamical process.   We show that in order to have a reasonable description of hadronization the local quark (gluon) density must not exceed a critical density calculated from the MIT bag model.  If the local density is larger than such a 
  value (which is equivalent to having a high temperature or energy density) the quarks cannot recombine to form hadrons (or the gluons decay into $q \bar q$ pair).  From a comparison of our results to LQCD we realize that in the QGP phase our $\epsilon/T^4$ is higher than the value suggested by
  lattice calculation and indeed it approaches the Stefan-Boltzmann limit as it should be.  Knowing that there is Debye color screening in the QGP phase we calculate the corresponding  mean field and adjust the parameters to reproduce lattice data.
    
 Quantum statistics (i.e. Pauli and Bose statistics) are  included
 similarly to   \cite{bert1}
 for Bose and \cite{sa04} for Fermi statistics\cite{tan}. 

 \section{Formalism}

 The mean free path method discussed above has been studied in detail at low energies and it has been
shown to solve the Boltzmann eq. in the cases
 where an analytical solution is known \cite{aldo}.  We have
generalized the approach to keep into account relativistic effects.  The particles move on straight lines
during collisions since we have not implemented any force yet.  For short we name the method proposed
 as Relativistic Boltzman equation (ReB).  In order to test our approach, we have discussed also some simple cases where analytical solutions are known to verify the sensitivity of our numerical approximations \cite{aldo,daimei,tan}.

In the present  calculations we  enclose a pion gas in a box with periodic boundary conditions.  The total energy of the system and the density is fixed for each calculation case according to the Bosonic nature of the system \cite{tan}.  
The temperature can be calculated at each time step (after equilibrium has been reached) from the momentum distribution of the particles.  The pions can collide elastically or inelastically according to the measured data.  Resonances
are included and their decay is simulated as well.   
\begin{figure}[ht]
\centerline{\hspace{-0.5in}
\includegraphics[width=4.5in,height=3.0in,angle=0]{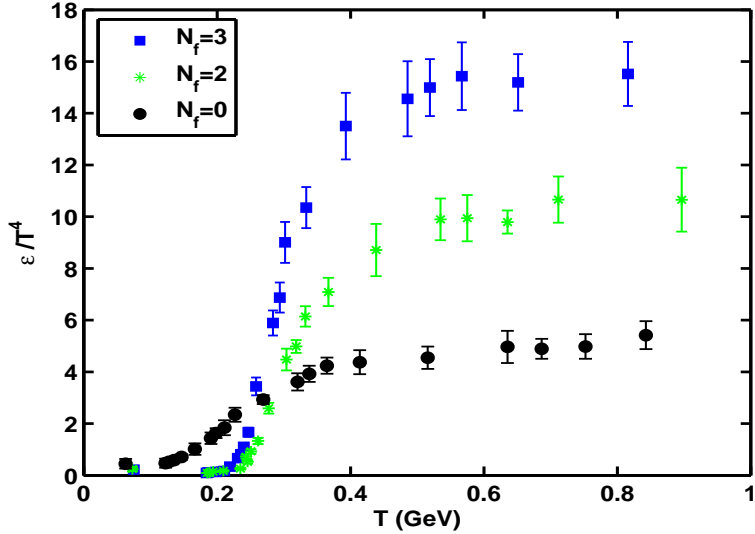}}
\vspace{0.2in} \caption{Energy density versus temperature for  a QGP with $N_f=$ 0 (full circles), 2(asterisks) and 3 (full squares) .  }
\label{fig1}
\end{figure}

We recall how we include a QGP in our approach.  First,   for a massless quark and gluon plasma in equilibrium the following relations hold for the pressure P, quark (antiquark and gluon) density $n_{q,(\bar q),(g)}$ and energy density $\epsilon$ versus temperature T\cite{las,wong,karsch,tan}:
\begin{equation}
P=g_{tot}\frac{\pi^2}{90}T^4;	
	n_q=n_{\bar q}=1.202\frac{3g_{q}}{4\pi^2} T^3;	
	n_g=1.202 \frac{g_g}{\pi^2}T^3;
\end{equation}

where $\epsilon=3P$, $g_{tot}=16+\frac{21}{2} N_f$, $N_f$ is the number of flavors.  In the MIT bag model\cite{las,wong} quarks and gluons are confined in the bag if the pressure is less than the critical pressure B that the bag can sustain.   Thus from the previous equation we can assume that the quarks and gluons are liberated in a collision if the energy density is larger then 3B.  This gives a critical energy density $\epsilon_c=3B=0.71 GeVfm^{-3}$ using $B^{1/4}=0.206 GeV$\cite{las,wong}.  For each h-h elementary collision we know  the energy of the  collision and the interaction  volume from the distance between the colliding particles in their center of mass system. This distance is taken as the radius of a sphere enclosing the two colliding particles and subsequently the newly formed partons. Thus we can calculate the number of quarks, antiquarks and gluons as a function of the energy density liberated in the collision inverting equation (1).
We stress that these relations are 
 strictly valid in thermal equilibrium thus it is perfectly justified in this work since we are mainly discussing equilibrium features of our system.  The partons are followed in time exactly like the hadrons solving the transport equation.  In particular the partons can collide elastically with other partons and hadrons using a cross section of 1 mb.  If in a collision between a parton and a hadron the local energy density is larger then the critical value, new partons are liberated from the hadron similarly to the mechanism discussed above.  The possibility of collisions among partons and hadrons is necessary in order that the total system can reach thermal equilibrium.  After a certain time $\tau$, the quarks can hadronize and the gluons can decay into
 a $q,\bar q$ pair which later can hadronize as well.  In the next  section we will discuss the case when the parameter $\tau$ is infinite i.e. once the partons are formed they do not hadronize.  In the following section we will discuss a 
 possible approach to hadronization.

\section{Equation of State Including a Mean FIeld}

The results of our calculation when including the QGP are displayed in fig.(1), for $N_f=0,2,3$, where we have used 10 MeV mass for the $u,d$ quarks and 160 MeV for $s$ quarks.  The relevant equation (1)  has been
 generalized to the finite quark masses used here. 
 
  In the figure the two main features of the system are seen, i.e. at low temperature the ratio 
$\frac{\epsilon}{T^4}$ is less than 1 which is the value expected for a mixture of bosons with masses \cite{las,wong}.  On the other hand, the ratio is  about 4, 12 and 16 as it should be for 0, 2 and 3 flavors quark system respectively. 
Such values are larger than the values  obtained in LQCD calculations \cite{karsch}.  This is evidently due to the neglect of an interaction among partons which is clearly important in LQCD.  On the other hand the good agreement with the expected high temperature limits strengthens our numerical calculations.

In our model  there is not a phase transition  but simply a crossover from an hadronic state   at low T to a state of partons at high T.  However, the crossover is rather visible and it should have some effect in experimental data at RHIC such as large fluctuations of D-mesons for instance \cite{terr05,terr04}.
  We obtain a cross over above 175 MeV  temperature depending  on the number of flavors .  These features are  at least in qualitative
agreement with microscopic LQCD results\cite{karsch} which strengthens our model somewhat. In fact, since the EOS of the two systems is similar, they should give similar consequences when features of heavy ion collisions are investigated.  We stress that our approach being a kinetic one could be easily extended and applied to relativistic heavy ion collisions which is the main purpose of this work, even though we will have to pay a particular attention to hadron-hadron phenomenology at higher momentum transfers.  Thus within this model we could study features of the collision with and without the phase transition and compare to data and/or be of guidance to more experimental investigations.
\begin{figure}[ht]
\centerline{\hspace{-0.5in}
\includegraphics[width=6.5in,height=4.0in,angle=0]{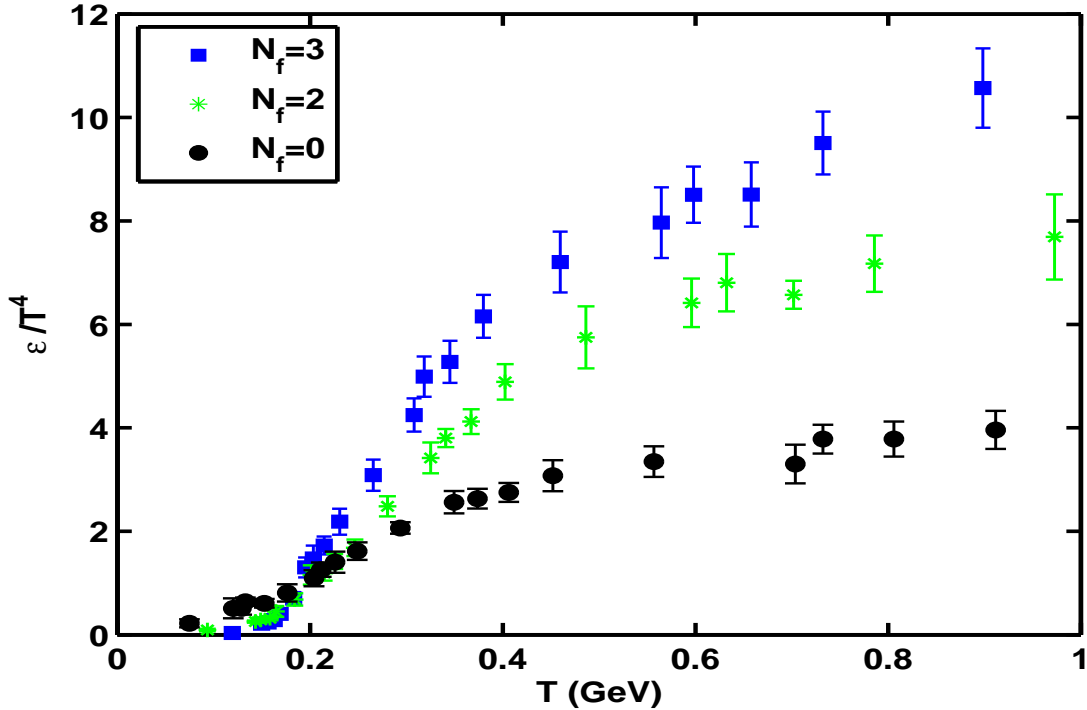}}
\vspace{0.2in} \caption{Same as in fig.1 but with the mean field included for each $N_f$ in the QGP phase.}
\label{fig2}
\end{figure}

In order to improve the results displayed in fig.(1) and obtain a better agreement to the LQCD calculations, we recall that in the QGP phase the linear potential which
keeps the quarks confined disappears while the Coulomb-like term due to one gluon exchange is screened due to the color charges and similar to the Quantum ElectroDynamics (QED) case.
In particular it is easy to realize that in order to reduce the ratio $\frac{\epsilon}{T^4}$ we need to introduce an attractive mean field $U_m$ which increases the kinetic energy of the partons, thus the temperature.  We can find 
the value of $U_m$ for the different flavors by fitting to the LQCD results. In fig.(2) we plot the EOS obtained including the mean field. Clearly, the values of $\epsilon/T^4$ at high T are in agreement with the LQCD results\cite{karsch}.  

The resulting mean field $U_m$ is plotted in fig.(3) as function of T.  A clear linear dependence on T is observed.  
\begin{figure}[ht]
\centerline{\hspace{-0.5in}
\includegraphics[width=6.5in,height=4.0in,angle=0]{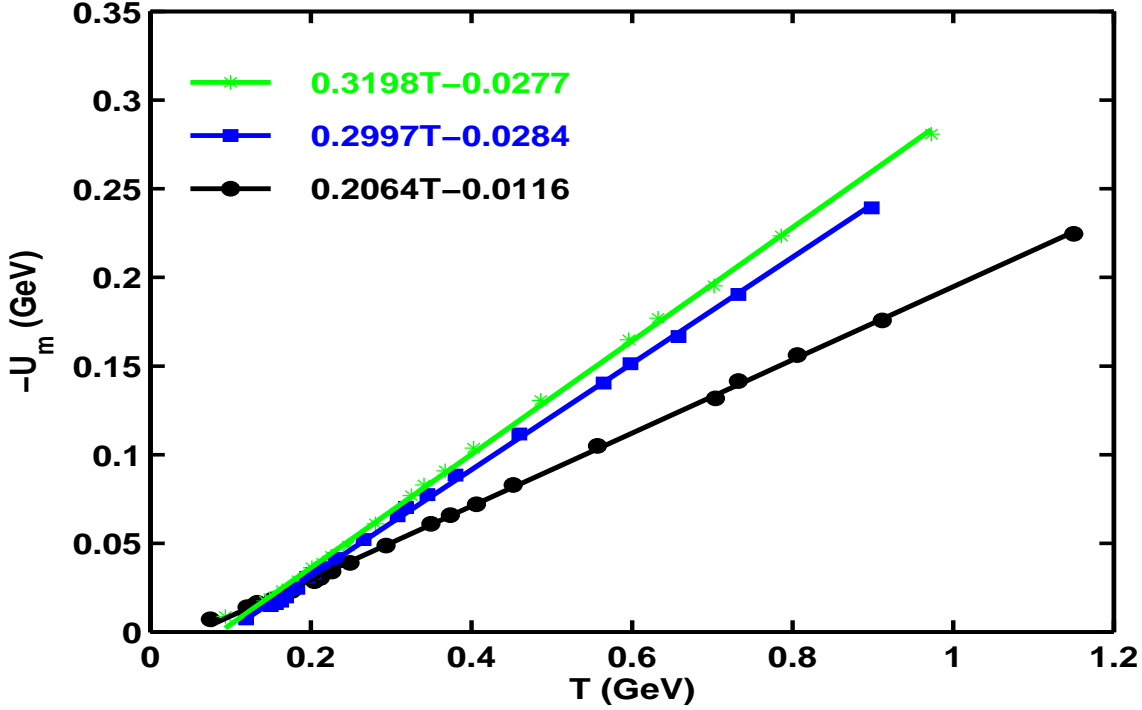}}
\vspace{0.2in} \caption{Mean field  vs T for each $N_f$ , symbols as in fig.(1). The best fit values to $U_m$ are given in the figure.}
\label{fig3}
\end{figure}
We can easily understand this results in the Abelian limit.  Consider for instance a quark  located at r=0.  This quark will interact with another quark or antiquark  located  at a distance r through a screened Coulomb potential\cite{wong}:
\begin{equation}
V(r)=-\frac{4}{3}\alpha_s \frac{e^{-m_d r}}{r}
\end{equation}
where $m_d$ is the screening mass\cite{wong} and  $\alpha_s$ is the strong coupling constant.  The mean field can be easily obtained summing up over all possible charges and integrating over the total volume at constant density:
\begin{equation}
U_m=\int 4\pi r^2 \rho V(r)dr=- \frac{4 \pi \alpha_s\rho}{m_d^2}
\end{equation}
Comparing eq.(3) to the figure(3) one clearly finds that $m_d\propto T$, since the density $\rho \propto T^3$, which is a well known result\cite{wong}.  
In particular knowing that in the Abelian approximation:
\begin{equation}
m_d^2(Ab)=\frac{g_q q^2T^2}{6}
\end{equation}
with the charge  $q^2=\frac{4}{3} 4\pi \alpha_s$.  For $N_f>0$, we can derive the value of
$m_d$ , eq.(3), using the fit values in fig.3 and we can compare the result to eq.(4) obtaining:   $m_d=1.9 m_d(Ab)$, i.e. our estimate based on
a fit to LQCD results, is about a factor two larger than the Abelian limit, eq.(4).  This difference could be expected since the gluons which have been neglected in the Abelian limit play a role as well and their dynamics is also modified by some force as we have seen in figs.(2) and (3)  for $N_f=0$. 

\section{a simple scenario for hadronization}
\begin{figure}[ht]
\centerline{\hspace{-0.5in}
\includegraphics[width=6.5in,height=4.0in,angle=0]{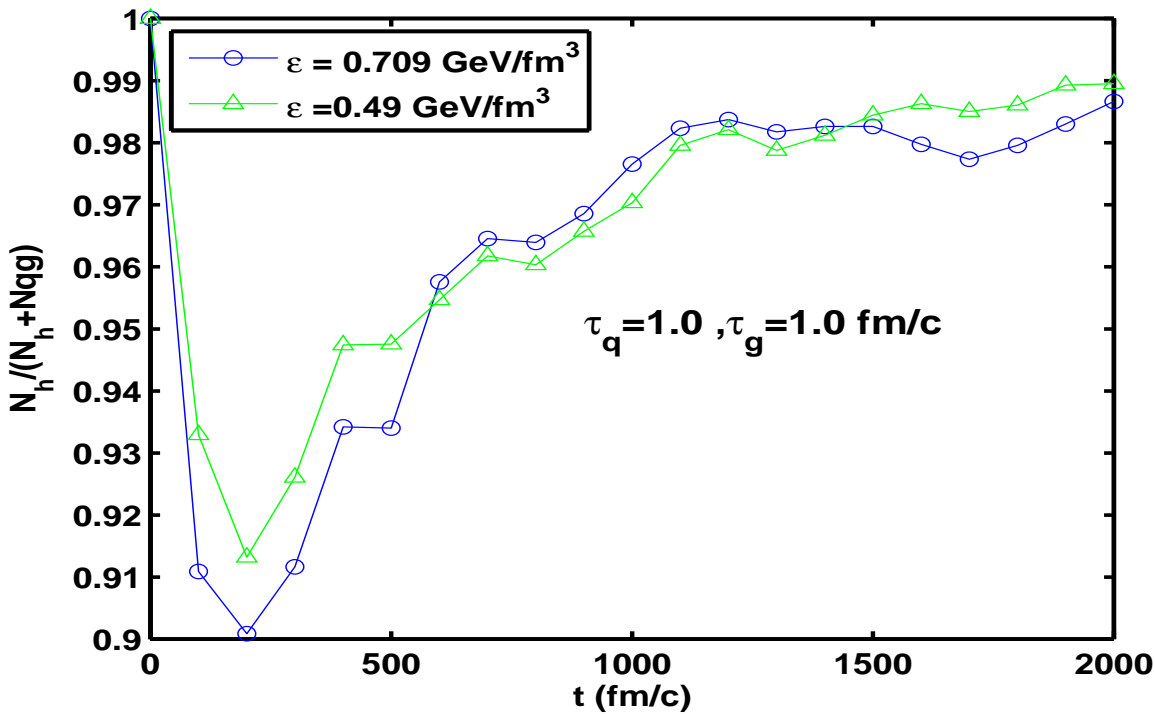}}
\vspace{0.2in} \caption{Number of hadrons divided the total number of particles versus time at two different energy densities.}
\label{fig4}
\end{figure}
In the previous section we have discussed the EOS assuming that after the partons are created they do not recombine again to form new hadrons.  
Of course we expect that this is true for the box calculations at very high temperatures, i.e. in the QGP phase.  However at temperatures near the
cross over and even in the hadronic phase it might happen that because of fluctuations in a given region the energy density is larger than the
critical value and partons can be created.  Since the energy density is on average not so large there might be a possibility for partons to
combine again.  We would like to discuss here some possible ways to implement this mechanism and check how the EOS is modified.  Already in a 
previous work of simulations of heavy ion collisions we had introduced a finite lifetime $\tau$ for quarks to combine into hadrons \cite{daimei}. To be
 more precise we let the quarks created at a given time $t$ evolve to a time $t+\tau_q$ (calculated in the quarks rest frame) before they can hadronize, while gluons decay after a time  $t+\tau_g$ into a $q\bar q$ pair which can eventually hadronize after a time  $\tau_q$. 
 For the hadronization we assume that quarks can combine to form resonances depending on their total energy.  The use of resonances is important
  to conserve energy and total momentum.
  Before hadronization the partons move on straight line trajectories and they can only collide elastically.  If we adopt this scenario we obtain
  a surprising (at first) result which is plotted in fig.(4). 
 In the figure the total number of hadrons divided the total number of particles (hadrons plus
  partons) is plotted versus time for two different energy densities using a formation time of 1.0 fm/c for quarks and a similar value for the decay time of gluons into  $q \bar q$.  At these energy densities from the previous figure (1) we expect to be at very high temperature i.e. in the QGP phase. But 
  as we see from the figure (4), no matter how large the energy density is the system hadronizes completely after a few hundreds of fm/c.  Thus the
  EOS that we get with this assumption of a finite formation time is similar to the Hagedorn one.  In fact, after the first collisions some partons are 
  created which later on hadronize in a larger number of hadrons than initially.  Because of the increase of the number of hadrons, the temperature 
 decreases and  eventually saturates to a value of the order of the pion mass (the Hagedorn temperature) even if we increase largely the energy densities.  
  \begin{figure}[ht]
\centerline{\hspace{-0.5in}
\includegraphics[width=6.5in,height=4.0in,angle=0]{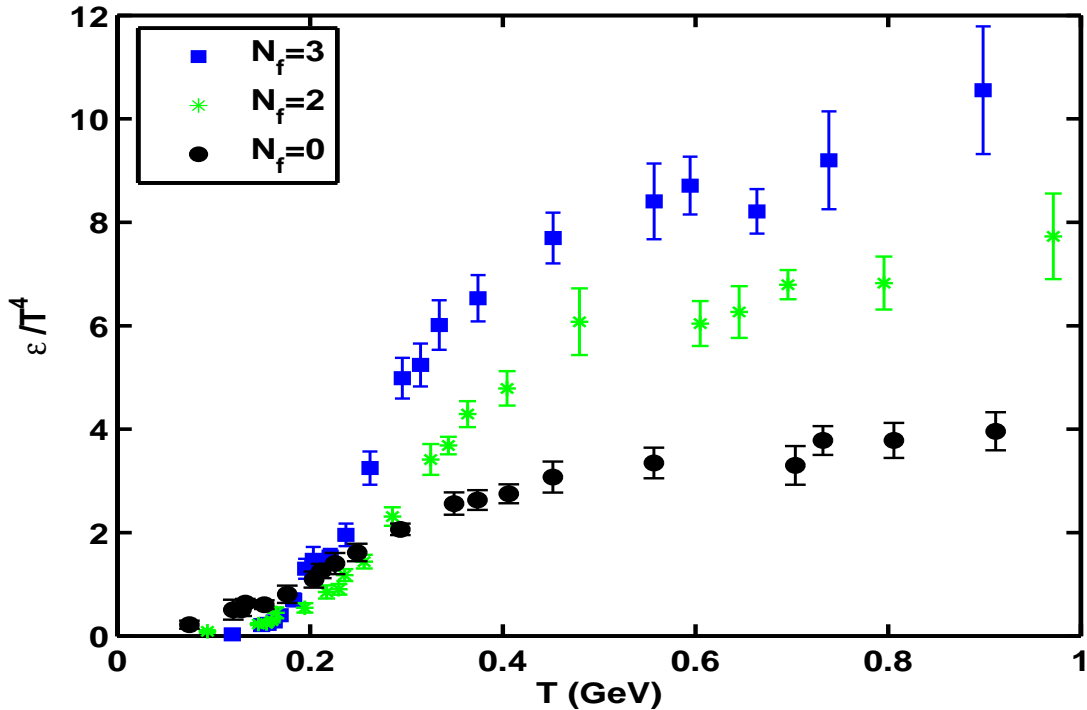}}
\vspace{0.2in} \caption{EOS including the mean field and with recombination times $\tau_q=1.0 fm/c$ and $\tau_g=0.5 fm/c$.}
\label{fig5}
\end{figure}

  In order to overcome this problem we can get some more insight in the hadronization mechanism using the Bag model again.  In the model the
  partons are confined when the density (local) is smaller than a critical value which can be obtained from eq.(1) for a fixed value of the bag constant
  and for each $N_f$ flavor.  Thus at each time step and for each particle we check the local density and if such a density 
  is smaller than the critical value, the $q\bar q$ pair  hadronizes or a gluon will decay into a $q\bar q$.

The results of the calculations with mean field included are given in figure (5), due to the inclusion of the condition on the critical density the partons 
cannot recombine at high T and the expected EOS is recovered.  Actually, the use of a condition on the density makes the use of  the lifetime 
somewhat unnecessary. In fact even if we put the recombination time equal to zero, the partons cannot hadronize if the density is too large.  

\section{conclusions}

In this work we have applied a recently introduced transport approach to
study the equation of state of a meson gas and its possible transitions to a Quark Gluon Plasma.
The model includes the possibility of resonance formation and  decay.  The possibility of a QGP
may be included based on the MIT bag model.  Quantum statistics can be easily included but we have 
seen that at the temperature discussed here those effects are negligible.  The Hagedorn limiting 
temperature is recovered for a pure hadronic system.  When including a QGP the obtained equation
of state is in qualitative agreement with LQCD calculations.  A better description could be obtained 
by introducing some interaction (attractive) among partons. We have shown that the mean field that we 
have derived to fit the LQCD results gives a Debye screening mass about a factor two larger than expected in the Abelian 
limit putting in evidence the role of the gluons in our model. We have also discussed a possible scenario for hadronization assuming that
$q\bar q$ could combine into hadrons or gluons split into $q\bar q$ pairs if the local density is smaller than a critical value derived in the 
Bag model.  Because of the condition on the density the particles are relatively close enough in space thus it is easy to combine into resonances
in order to conserve total energy and momentum of the system.  This, we feel, will be especially important when dealing with finite systems in
order to avoid having isolated quarks in phase space.

   The final goal of this work is to apply the model to relativistic heavy ion collisions at RHIC
and Cern energies. To this purpose we believe we have a better  control now on  the ingredients of the model
such as the effectively implemented equation of state.  Next we are going to concentrate on a   good reproduction of the available
data on elementary hadron-hadron collisions in order to be able to investigate heavy ion collisions. Our efforts in these directions will be discussed in 
following papers.

Finally, Z.G.T: acknowledge the financial support from  INFN and
Department of Phys. University of Catania in Italy (where most of
the work was performed) .  This work is supported in part
 by the EU under contract CN/ASIA-LINK/008(094-791).  


\begin{thebibliography}{99}
\bibitem{karsch}F. Kartsch, Nucl.Phys. A698, 199c(2002).
\bibitem{gei}
K. Geiger and B. M$\ddot{u}$ller, Nucl. Phys. B 369, 600 (1992);
K. Geiger, Phys. Rep. 258, 237 (1995).
\bibitem{aich}
S.Bass, et al., Prog.Part.Nucl.Phys. 41, 255 (1998).
\bibitem{cug}
J. Cugnon, T. Mitzulani, J. Meulen, Nucl. Phys. A 352, 505 (1981).
\bibitem{bert}
G. F. Bertsch and S. Das Gupta, Phys. Rep. 160, 189 (1988).
\bibitem{sa1}
Ben-Hao Sa, Wang Rui-Hong, Zhang Xiao-Ze, Zheng Yu-Ming, and Lu Zhong-Dao,
Phys. Rev. C 50, 2614 (1994).
\bibitem{aldo}
A. Bonasera, F. Gulminelli, and J. Molitoris, Phys. Rep. 243,1(1994); A.Bonasera, T.Maruyama Progr.Theor.Phys. 90(1993)12.
\bibitem{daimei} D.M.Zhou, S.Terranova and A.Bonasera,  nucl-th/0501083.
\bibitem{tan}Z.G.Tan, D.M.Zhou, S.Terranova and A.Bonasera,  nucl-th/0606055, and proc. XXII Winter Workshop on
Nuclear Dynamics, pag.31, W.Bauer et al. eds.(2006).
\bibitem{las}
L. P. Csernai, "Introduction to relativistic heavy ion
collisions", John Wiley and Sons, 1994.
\bibitem{wong}
 C.~Y.~Wong, {\it Introduction to High-Energy Heavy ion Collisions},
 World Scientific Co., Singapore,1994.
\bibitem{bert1}
G. M. Welke and G. F. Bertsch, Phys. Rev. C 45, 1403 (1992).
\bibitem{sa04} Ben Hao Sa and A. Bonasera, Phys. Rev. C 70, 034904 (2004).
\bibitem{terr05} S.Terranova, D.M. Zhou and A.Bonasera,
nucl-th/0412031, EPJA26, 333 (2005).
\bibitem{terr04} S.Terranova and A.Bonasera, Phys.Rev.C70,024906
(2004); T. Maruyama and T. Hatsuda, Phys. Rev. C 61, 62201(R) 
(2000).
\end{thebibliography}
\end{document}